\documentclass{osa-article}

\journal{oe}
\begin{document}

\title{Asymmetric high energy dual optical parametric amplifier for parametric processes and waveform synthesis}

\author{Brandin Davis, Tobias Saule, and Carlos A. Trallero-Herrero\authormark{*}}

\address{Department of Physics, University of Connecticut,
Storrs, Connecticut 06269, USA \\}

\email{\authormark{*}carlos.trallero@uconn.edu}

\begin{abstract}
We report on an asymmetric high energy dual optical parametric amplifier (OPA) which is capable of having either the idlers, signals, or depleted pumps, relatively phase locked at commensurate or incommensurate wavelengths. Idlers and signals can be locked on the order of 200~mrad rms or better, corresponding to a 212~as jitter at $\lambda$=2~$\mu$m. The high energy arm of the OPA outputs a combined 3.5~mJ of signal and idler, while the low energy arm outputs 1.5~mJ, with the entire system being pumped with a 1~kHz, 18~mJ Ti:Sapphire laser. Both arms are independently tunable from 1080~nm-2600~nm. The combination of relative phase locking, high output power and peak intensity, and large tunability makes our OPA an ideal tool for use in difference frequency generation (DFG) in the strong pump regime, and for high peak field waveform synthesis in the near-infrared. To demonstrate this ability we generate terahertz radiation through two color waveform synthesis in air plasma and show the influence of the relative phase on the generated terahertz intensity. The ability to phase lock multiple incommensurate wavelengths at high energies opens the door to a multitude of possibilities of strong pump DFG and waveform synthesis.
\end{abstract}

\section{Introduction}
The need for novel ultrafast laser sources at wavelengths far from the visible continues to be of extreme importance as new fields of study are developed. As an example, interest in ultrafast laser sources in the mid infrarad (MIR) to long wavelength infrared (LWIR) has grown substantially due to the applications these wavelengths provide in the molecular fingerprint region, including studies on vibrationally mediated photochemistry\cite{Delor2014}, infrared chemical nano-imaging\cite{Muller2015}, coherent control of vibrational dynamics\cite{Picon2011}, and femtochemistry\cite{martin2004femtochemistry}. Producing ultrafast infrared pulses capable of strong field studies is still an ongoing endeavor\cite{Wilson2019,Cheng2020,Rossi2020}, with one of the first examples of strong field ionization of a noble gas in the LWIR being shown only recently\cite{Wilson2019}. Furthermore, the need for ultra-fast pulses beyond 20~$\mu$m has grown in recent decades for strong interactions with quantum materials\cite{Matsunaga2013,Yang2019,Vaswani2020}. Even more generally the synthesis of light transients of multiple octaves \cite{Chan2011Science, Wirth2011Science, Huang2011NAtPhot, Hassan2012RSI} holds many promises, such as the generation of intense table top attosecond pulses \cite{Tosa2012, Jin2014, Jin2016, Xue2020ScienceAdv} or the control and measurement of phase transitions in materials \cite{Schiffrin2013Nature}. However, the generation and control of multiple incommensurate waveforms is also of great interest as it can lead to waveform synthesis of multiple octaves \cite{osti_1188316,PhysRevA.102.013520,Liang2017}, attosecond XUV pulses \cite{Dao2015NatComm} and tunable IR pulse generation from the LWIR to THz \cite{Balciunas2015} through electronic currents generated in air \cite{Hamster1993,Kim2008,Kim2007,Koulouklidis2020}. With this method, THz pulses as short as a single cycle\cite{Clerici2013,7094213} and pulse energies as high as 0.185~mJ with energy conversion efficiencies of 2.85\%\cite{Koulouklidis2020} and tunability from the MIR to THz\cite{Balciunas2015} have been shown. To take advantage of these techniques there is need for flexible, high pulse energy, wavelength tunable laser sources that output multiple beams which can be individually tuned and have their relative phases locked. Dual-OPAs as a source are an ideal candidate, owing to their large wavelength tunability, energy scalability, and pairs of outputs which can be individually tuned. While advantageous, few studies have been done on the experimental feasibility of high energy dual-OPAs, one of the most extensive of which being published in 2020\cite{Rossi2020}. That system represents the state-of-the-art for dual-OPAs and it showcases their ability to perform waveform synthesis. However, such a system is based upon an extensive custom design which additionally employs a carrier envelope phase (CEP) stable white light source and the total output power of which is limited to sub-millijoule. This rather low energy limits non-linear processes, like LWIR and THz generation techniques, to a power regime which make high intensity studies challenging.

To preserve the flexibility, increase the power, and decrease the complexity of such a system, we test the feasibility of phase sensitive applications with a commercially available dual-OPA platform without the requirement of a CEP stable seed. A readily available platform in reach of the broader scientific community will have a large impact on the design of future systems.

In this article we demonstrate a high energy, asymmetric, dual-OPA capable of generating multiple millijoules of energy per pulse at 1~kHz repetition rate. We further show that both the signals and idlers can be relatively phase-locked, thus producing up to six highly energetic beams that are pairwise phase stable with only the use of a spectrometer and piezo based stage for feedback. The presence of asymmetry in the energy per pulse allows for a great variety of parametric processes in the strong-pump regime such as DFG. First proof of principle of waveform synthesis with our dual-OPA is demonstrated through generation of THz radiation in an air plasma.

\section{Experiment}
\begin{figure}[h!]
\centering\includegraphics[width=\textwidth]{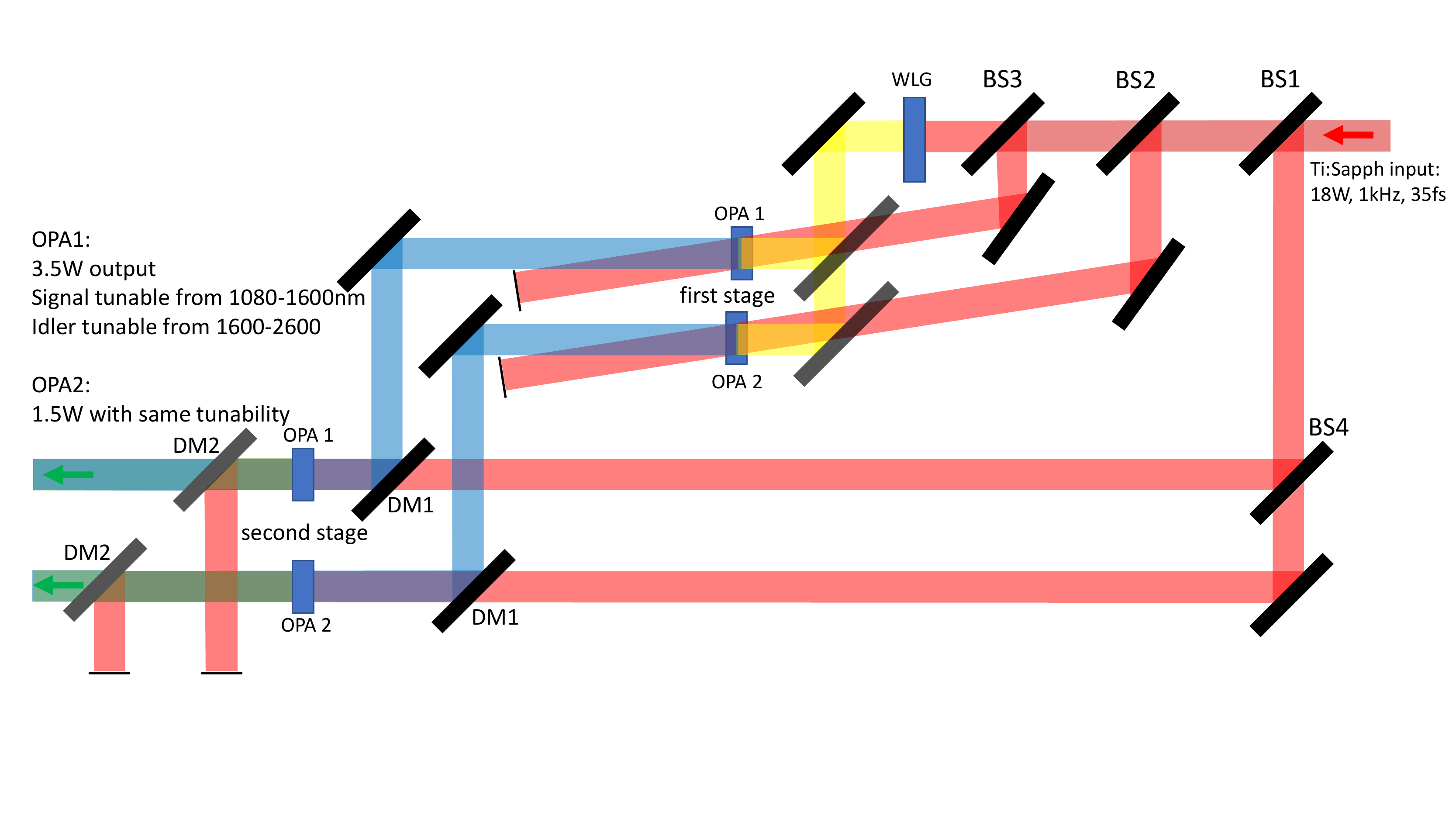}
\caption{Schematic of the dual-OPA. Pump beam is passed through BS1 (R=97.5\%) and the reflected light through BS4 (R=70\%) providing the energy for each of the final OPA stages. Light transmitted through BS1 encounters BS2 (R=37.5\%) and BS3 (R=95\%), which provide pump power for the 1st stage non-collinear OPAs. The remaining power is focused into a sapphire plate used for white light generation (WLG) after which it's split in half and used to seed the first stage OPAs. Amplified first stage signal light is then collinearly overlapped with the second stage pumps to achieve the final outputs: OPA 1: 3.5~W tunable from 1080-2600~nm, OPA 2: 1.5~W tunable from 1080-2600~nm. Powers are for combined signal and idler.}
\label{fig:Figure 1}
\end{figure}

Our experimental setup starts with a Ti:Sapph laser from Continuum USA, capable of delivering 18~mJ of energy per pulse, 35~fs in duration, at 1~kHz repetition rate, and 800~nm center wavelength. The amplifier is seeded by a commercially available Ti:Sapph oscillator from Laser Quantum, producing over 560~mW of average power at 80~MHz repetition rate with a pulse duration of under 8~fs. This seed is stretched to over 1~ns, and injected into a regenerative amplifier where the repetition rate is reduced to 1~kHz. Pulses are subsequently amplified in two water cooled Ti:Sapph multipasses followed by a cryogenically cooled multipass and then compressed with a grating compressor pair. The output beam pointing is stabilized with a TEM Aligna system used before the compressor gratings. Before amplification, pulses are sent through a Dazzler and Mazzler from FASTLITE, allowing for control of the spectral phase and amplitude of the output pulses. Higher orders of phase are eliminated through the use of the Dazzler, and pulse durations as short as 16~fs straight out of the compressor are possible through the combined use of the Dazzler and Mazzler.
The OPA is a modified dual-OPA design from Continuum/Quantronix composed of two OPAs, each comprising two stages based on bismuth triborate (BIBO) crystals cut for type II phase matching and sharing the same white light seed. The high energy arm of the OPA will be referred to as OPA 1 and the low energy arm as OPA 2. Figure \ref{fig:Figure 1} shows a schematic of the dual-OPA layout. 
The first beamsplitter (BS) reflects 97.5\% of the input which is then further split with a 70\% reflective BS to provide the two asymmetric pump beams which will be used in the final stage BIBO crystals. The power transmitted through the first beamsplitter is passed through two additional beamsplitters, BS2 being 37.5\% reflective and BS3 95\% reflective, to provide power for the pumps of both first stage OPAs. The remaining power is used to generate the white light seed via super continuum generation in a sapphire plate. This white light is split in half and used to non-collinearly seed both sets of first stage OPAs. Around 6~mW of signal light is produced in each of the first non-collinear OPAs (NOPA) which are then expanded and collimated to be amplified collinearly in the final stage of each arm. Each arm is capable of producing a signal with wavelength tunability from 1080-1600~nm and pulse durations as short as 50~fs (measured through second harmonic generation frequency resolved optical gating  (SHG-FROG)\cite{199311}), and an idler tunable from 1600-2600~nm with pulse durations as short as 100~fs. OPA 1 at maximum efficiency (~30\%) outputs 3.5~W of combined signal and idler, similarly OPA 2 produces 1.5~W.

\begin{figure}[h!]
\centering\includegraphics[width=\textwidth]{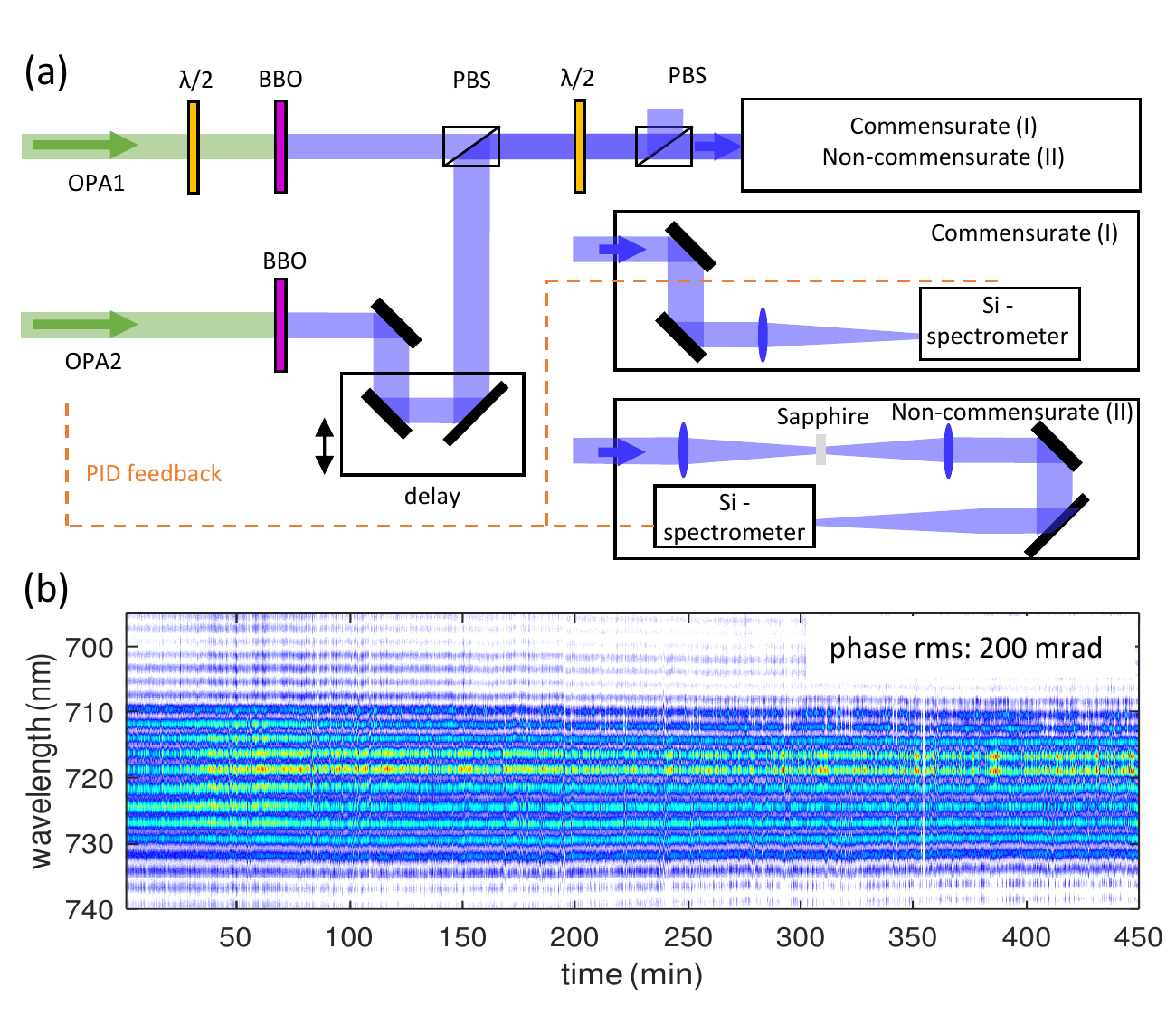}
\caption{(a) Relative phase measurement setup. Pairs of idlers or signals from OPA 1 and 2 are sent into separate arms where one has its polarization rotated with a half-waveplate ($\lambda$/2) to allow for overlap in the polarizing beam cubes (PBS) and each are frequency doubled in beta barium borate (BBO) crystals. Depending on the wavelengths of the inputs they are sent directly into a spectrometer (I), or broadened in a sapphire plate first (II). Subsequent spectral fringes are then locked through a Fourier transform PID loop. (b) An example of single shot spectral fringes of the frequency doubled signals (both signals set to 1440~nm) locked over a period of 7.5 hours. The plot was generated using every 10th recorded spectrum. }
\label{fig:figure2}
\end{figure}

It is known \cite{Rossi2018} that the CEP of the signal, $\phi_{s}$ and idler, $\phi_{i}$ of an OPA can be given by $\phi_{s}^{CEP} = -\omega_{seed}*T + \phi_{p}^{AP}$ and $\phi_{i}^{CEP} = \omega_{seed}*T + \frac{\pi}{2}$ respectively, where $T$ is the relative arrival time difference between the seed and the pump, $\phi_{p}^{AP}$ is the absolute phase of the pump, and p,s,i stand for the pump, signal, and idler respectively. As can be seen, the idler of an OPA is passively CEP stable while the signal has the CEP value of the pump. Assuming the white light seed was split in two, a second OPA could be seeded with CEPs $\phi_{s2}^{CEP} = -\omega_{seed}*T_2 + \phi_{p2}^{AP}$ and $\phi_{i2}^{CEP} = \omega_{seed}*T_2 + \frac{\pi}{2}$. Examining the respective signal and idler CEP relations, a relative phase for the signal and idler can be expressed as $\Delta\phi_s = \phi_{s1}^{CEP} - \phi_{s2}^{CEP}= \omega_{seed}(T_2 - T_1)$ and similarly $\Delta\phi_i = \omega_{seed}(T_1 - T_2)$. Therefore each pair of signals and idlers of a dual-OPA are each passively relative phase locked, with the idlers having the added benefit of being passively CEP locked. Having a well defined relative phase value is required in applications of waveform synthesis where enhancements quickly diminish as the relative phase between the electric fields changes. Some of these applications include increased yield and cutoff of high harmonic generation and air plasma THz generation \cite{Jin2016,Zhang2016}.The experimental setup is shown in Figure \ref{fig:figure2}(a), each beam of either the signals or idlers is sent into separate arms of the relative phase setup where both are frequency doubled to allow the use of silicon based spectrometers. 
Their second harmonics are recombined using a polarizing beam splitter (PBS) cube and a half-wave plate. Using a second PBS and half-wave plate pair their polarization's are overlapped and relative intensities are adjustable. If the input wavelengths are the same, they are directly focused into a spectrometer where spectral fringes are obtained as seen in Figure \ref{fig:figure2}(b). 
\begin{figure}[h!]
\centering\includegraphics[width=\textwidth]{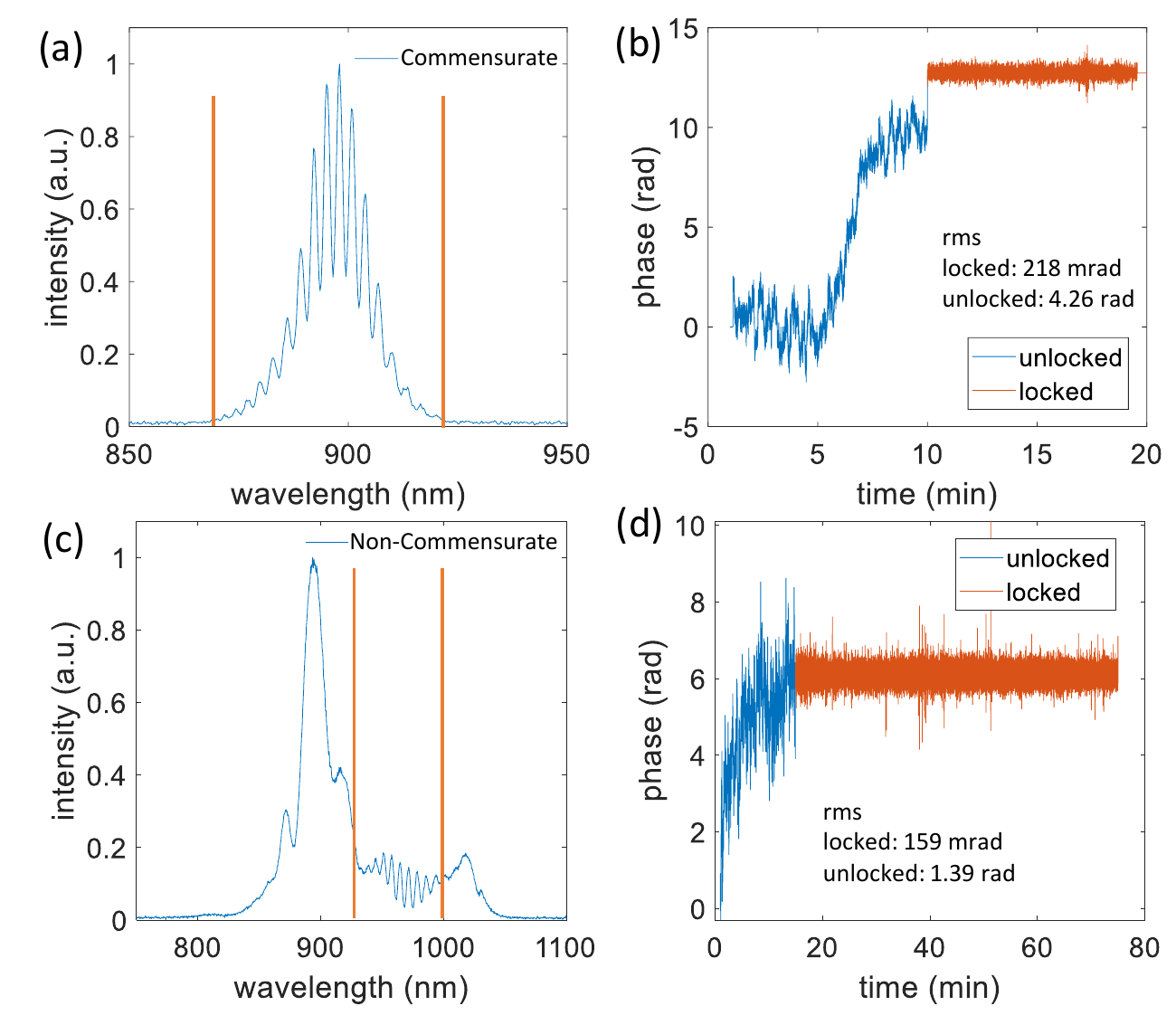}
\caption{(a) Spectral fringes and (b) relative phase values of the frequency-doubled idlers set to the same wavelength (1790~nm). (c),(d) Same as (a),(b) but for the incommensurate case (1790~nm and 2040~nm). Vertical red lines in (a), (c) denote area of interest for the Fourier transform used to calculate phase values. In (b), (d) red lines show relative phase with active locking compared to free running in blue.}
\label{fig:figure3}
\end{figure}
For non-commensurate wavelengths the beams are focused into a sapphire plate to allow for spectral broadening and subsequent spectral overlap. 
\section{Results}
To verify this assertion we measured the relative phase between the respective signals and idlers from our dual-OPA. The spectral fringes oscillate with a phase given by: $\varphi(\omega) = 2\phi_{(s,i)1} - 2\phi_{(s,i)2} + 2\omega_{(s,i)}\tau$, where $\phi_{(s,i)}$ is the spectral phase of the respective fundamental pulse, and $\tau$ is the time delay between the two pulses. Using this and the equations above, we can see that $\phi_{(s,i)1} - \phi_{(s,i)2} = \omega_{seed}(T_{2,1} - T_{1,2})$ , and therefore $\varphi(\omega) = 2\omega_{seed}(T_{2,1} - T_{1,2}) + 2\omega_{(s,i)}\tau$. For the fundamental value of the relative phase this would then be divided by two to account for the second harmonic having twice the phase of the fundamental. All future references to a relative rms phase value will take this into account. Since all variables are time independent ideally the relative phase should be locked. Due to experimental conditions there are variations in all three of these variables which cause short term jitter (stage position jitter, intensity fluctuations, air fluctuations, vibrations in the opto-mechanics of the optical path and small fluctuations of $\omega_{seed}$ etc) and long term drifts (thermal drifts, beam pointing, etc). To eliminate these long term drifts and reduce short term jitter a piezo driven stage was added in one arm of the setup and the spectral fringes were stabilized using a proportional–integral–derivative (PID) feedback loop to control the piezo, additionally the beam pointing of the Ti:Sapph laser is actively stabilized. The long term drifts and short term jitter can be seen in Figure \ref{fig:figure3} (b,d). Plotted is a direct comparison of the relative phase between the idlers while unlocked (blue) and locked (red), sampled at a frequency of 75~Hz and 4~ms integration time (the factor of two due to averaging over four pulses was included in the rms values). The width of the data corresponds to short term variations in relative phase while the drift from the starting value corresponds to the long term variations. While the short term variations are on the order of 200~mrad, the long term variations are quite significant, ranging from 1-10 radians in just ten minutes. Without feedback these long term drifts would make phase sensitive measurements impossible if long run times were required. The locked data shows that while the feedback loop is active, long term drifts are eliminated, leaving only short term variations. Figure \ref{fig:figure2}(b) demonstrates the longevity of the relative phase lock of the spectral fringes of the second harmonic of the signals, which were locked for over seven and a half hours with a relative phase rms value of 200~mrad. The lock time is only limited by the natural long term relative drift and therefore the range of the piezo. Not only can the relative phases be locked for extended periods of time, but with full control of the piezo, arbitrary relative phases can be chosen and locked to, giving full control of the relative phase of the two-color waveform.
\begin{figure}[ht]
\centering\includegraphics[width=0.9\textwidth]{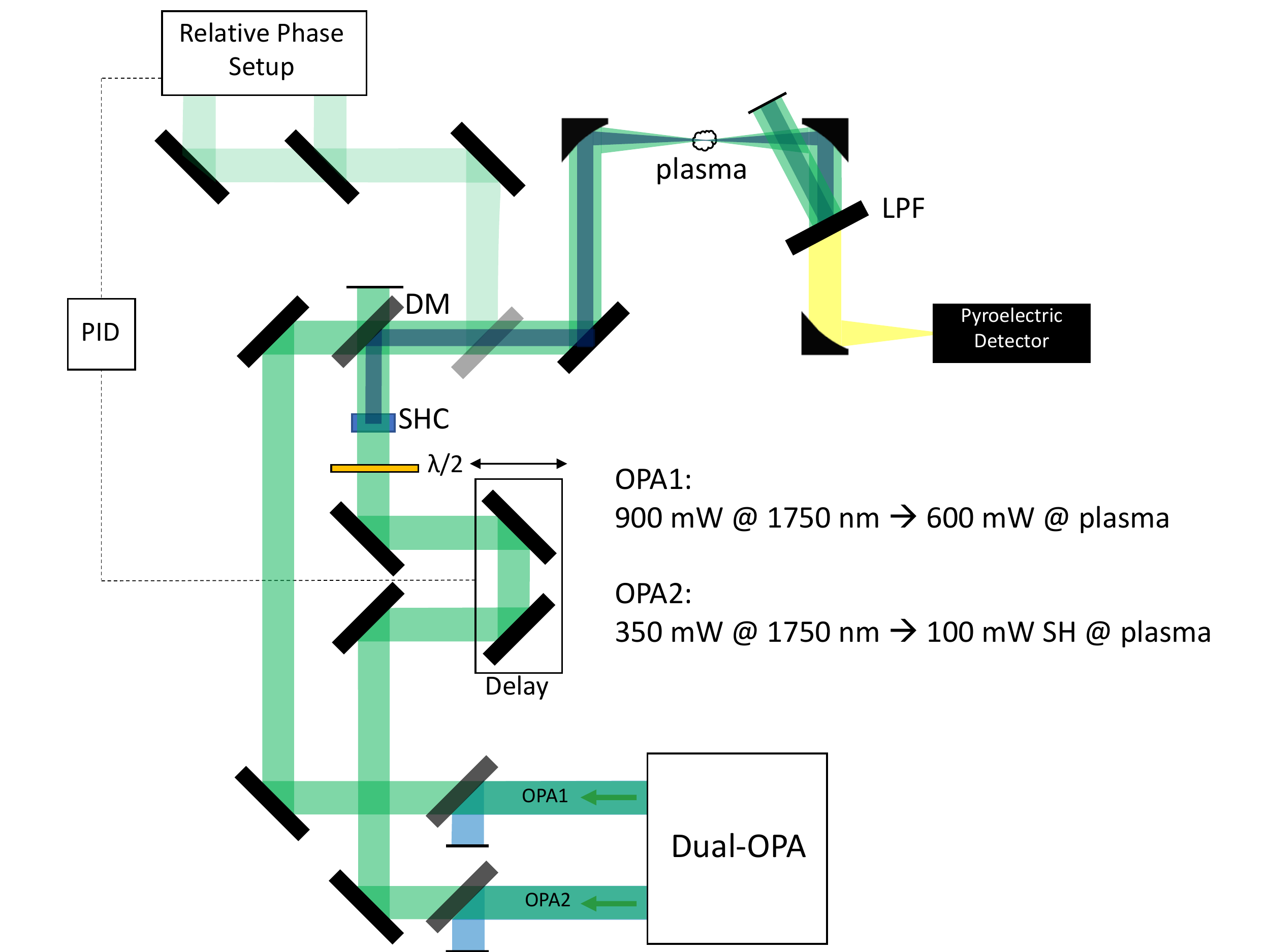}
\caption{THz generation setup. Idlers are sent into the separate arms of the setup where one is frequency doubled and recombined with the other. A small reflection of the waveform is taken with an uncoated CaF$_{2}$ wedge and then split using a 1500~nm long pass filter and sent into the relative phase setup to have their phases locked. The transmitted waveform is sent into a two inch focal length off-axis parabolic mirror and focused in ambient air. Generated THz pulses are sent through a 2.5~$\mu$m long pass filter and refocused onto a pyroelectric detector. SHC, second harmonic crystal; DM - dichroic mirror, LPF - long pass filter}
\label{fig:figure4}
\end{figure}

As a a demonstration of waveform synthesis, we generate THz radiation from air plasma \cite{Xie2006, Jang2019}. This technique for generating THz radiation has become a popular alternative to the traditional method of optical rectification in nonlinear crystals due to having no damage threshold, no absorption edges limiting gain bandwidth, and its scalability. It requires an asymmetric electric field, which is typically obtained by focusing the fundamental laser through a second harmonic generation crystal and adjusting the relative phase producing the asymmetric field. Due to the monolithic implementation of the method, the $\omega$ and 2$\omega$ fields are locked in their relative phase, which can be adjusted, typically by rotations of the second harmonic crystal.
\begin{figure}[ht]
\centering\includegraphics[width=\textwidth]{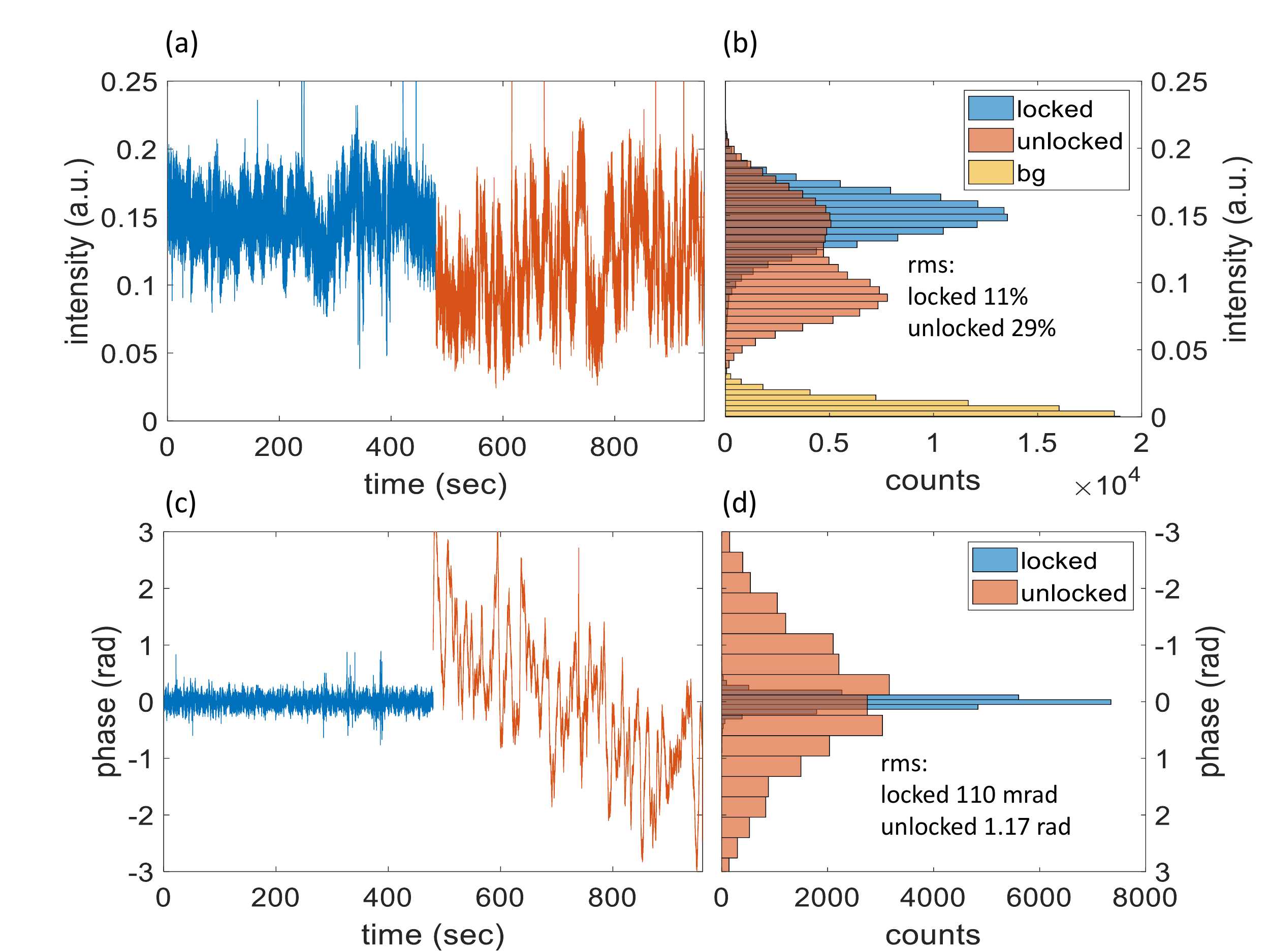}
\caption{(a) THz pulse energy as measured by a pyroelectric detector, with relative phases locked (blue) and unlocked(red). (b) Histogram of energy values from (a), with background counts shown in yellow. (c) Simultaneously measured relative phase of the two idlers whie locked (blue) and unlocked (red). (d) Histogram of phase values in (c) with locked and unlocked relative phases. Power measurements were taken shot-to-shot and phase measurements averaged over four pulses.}
\label{fig:figure5}
\end{figure}
The downside of this method is that typical SHG efficiency is on the order of 5\%, whereas for maximum field asymmetry, and thus maximum THz generation efficiency, the second harmonic intensity should be on the order of 50\% of the fundamental \cite{Zhang2016}. The monolithic design, while ensuring relative phase stability, doesn't allow for individual control of the respective beams, such as spot size, wavelength, or power tuning. In particular, being able to tune the wavelength ratios allows for the generation of ultra-fast light from the THz regime to the LWIR \cite{Balciunas2015}. With a dual-OPA every frequency ratio from 1 to 0.5 can be produced and relatively phase locked allowing for full tuning from THz to the LWIR. To show the effect of the relative phase on THz generation, we implemented a $\omega$-2$\omega$ THz air plasma generation setup, shown in Figure \ref{fig:figure4}. Both idlers of the dual-OPA were set to 1750~nm, the idler from OPA 2 has its polarization flipped such that its second harmonic has the same polarization as the idler from OPA 1. They are then combined to create the $\omega$-2$\omega$ asymmetric field. A small percentage of this field is picked off with an uncoated calcium fluoride wedge and sent into the relative phase setup. The beams passing through the wedge are aligned such that they are collinear and then focused into ambient air with a off axis parabolic mirror with an effective focal length of 2 inches, then collimated, and passed through a 2.5~$\mu$m long pass filter, and refocused onto a Molectron P-301 pyroelectric detector. To show the effect of the relative phase of the synthesized field on THz generation we measured the THz signal with the relative phase of the $\omega$-2$\omega$ field locked and unlocked, the results of which are shown in Figure \ref{fig:figure5}. When locked, the relative phase was scanned to find the point of highest THz signal on the pyroelectric detector. As can be seen, by locking the relative phase of the two beams the THz signal rms stability has a reduction in fluctuation from 29\% to 11\%. While unlocked the relative phase slipped by 4 radians, see Figure \ref{fig:figure5}(c). 

In addition, we performed experiments with incommensurate wavelengths, with one idler set to 1790~nm and the other frequency doubled to 1020~nm corresponding to a frequency ratio of 0.57, to produce a pulse at 7.3~$\mu$m center wavelength, not shown here. However, in that case we did not observe any influence of the relative phase in the generated power within our margin of error. We believe this is due to the fact that at incommensurate wavelengths with beat frequency periods much shorter than the pulse duration, scanning the relative phase doesn't change the shape of the waveform, it merely shifts in phase the produced waveform under the pulse envelope. Therefore regardless of the relative phase value, the part of the pulse responsible for ionization and THz production will see many cycles of the waveform being synthesized. This translates to a weak dependence on the output energy with the relative phase. On the other hand, at commensurate frequencies a shift in the relative phase completely alters the waveform being synthesized \cite{Kim2007}, leading to large changes in the output energy. While the relative phase for incommensurate frequencies will affect other characteristics of the generated THz beam, such as the CEP \cite{Balciunas2015}, the intensity should not be significantly altered with a change in the relative phase of the input beams.

\section{Conclusion}
To conclude, we have demonstrated the design and use of a high power, asymmetric, dual-OPA that is capable of producing pairs of signal and idler pulses that are individually tunable in wavelength and can be actively phase locked down to values of 110~mrad rms. Phase stability is accomplished pairwise between the two idlers, the two signals, or the two depleted pumps. For the idlers, at 2~$\mu$m center wavelength, the rms phase noise is equivalent to control at the 200 attosecond level. In addition to phase locking, which provides stability for the synthesized waveform, arbitrary relative phase values can be chosen, allowing for fine control of the waveform. This control and stability was demonstrated by producing THz radiation in air with the two idlers of the dual-OPA. Rms power fluctuations of the generated THz radiation was reduced from 29\% to 11\% after locking of the relative phases of the idlers. The ability to stabilize and precisely control the relative phase, and adjust the wavelength ratio, power ratio, and spot size of the two beams to be synthesized allows for a large amount of flexibility in optimizing the generation of beams with wavelengths from the XUV to THz regimes. Additionally, the high power asymmetric nature of this dual-OPA is ideally suited for driving parametric down conversion processes in the strong pump regime, enabling strong field studies in the MIR and LWIR. The number of laser sources capable of putting to use the full power of these techniques is limited, and are often more complex. With the ease of use and simplicity of implementing a relative phase locking setup, the unmatched output pulse energy and average power, all while being built on a commercial platform, makes our source a unique platform that will allow the broader scientific community to generate fully tunable synthesized waveforms and drive strong field parametric down conversion processes. It's interesting to note that if used with a CEP stabilized front-end laser, the dual-OPA would output six CEP stable beams, four of which are wavelength tunable, and two being the depleted pumps which themselves still have multi-millijoule levels of energy, increasing the wavelength combinations for DFG and the possibilities for waveform synthesis significantly.

\section*{Funding}
This research was performed under the Office of Naval Research, Directed Energy Ultra-Short Pulse Laser Division grant N00014-19-1-2339. T.S. was partially funded by the US Department of Energy, Office of Science, Chemical Sciences,
Geosciences, \& Biosciences Division grant DE-SC0019098.

\section*{Disclosures}
The authors declare no conflicts of interest.

\bibliography{References}

\end{document}